\documentclass{webofc}
\usepackage[varg]{txfonts}   
\begin{document}
\title{A Ceph S3 Object Data Store for HEP}

\author{
    \firstname{Nick} \lastname{Smith}\inst{1}\fnsep\thanks{\email{ncsmith@fnal.gov}} \and
    \firstname{Bo} \lastname{Jayatilaka}\inst{1} \and
    \firstname{David} \lastname{Mason}\inst{1} \and
    \firstname{Oliver} \lastname{Gutsche}\inst{1} \and
    \firstname{Alison} \lastname{Peisker}\inst{1} \and
    \firstname{Robert} \lastname{Illingworth}\inst{1} \and
    \firstname{Chris} \lastname{Jones}\inst{1},
    on behalf of the CMS Collaboration
}

\institute{
    Fermi National Accelerator Laboratory, Batavia, IL
}

\abstract{ We present a novel data format design that obviates the need for
data tiers by storing individual event data products in column objects. The
objects are stored and retrieved through Ceph S3 technology, with a layout
designed to minimize metadata volume and maximize data processing parallelism.
Performance benchmarks of data storage and retrieval are presented. }
\maketitle
\section{Introduction}
\label{intro}

The HL-LHC  poses a significant challenge for the CMS experiment's data
management model. A combination of higher pile-up and finer detector
granularity will result in each collision event of data or simulation requiring
considerably larger byte-storage~\cite{Software:2751565}. The overall event
rate to storage, currently at approximately 1 kHz, is also anticipated to
increase to at least 7.5 kHz. In the early years of Run 4, several new
subdetector components will come online, and their low-level data will need to
be validated and synthesized into high-level reduced data suitable for
analysis. This process will be challenging if the data products pertaining to
those detector elements are not easily accessible. Future innovative analysis
may necessitate different data reduction strategies. Given these
considerations, the amount of data that is regularly accessed by end-user
analysts is expected to grow significantly in the coming years.

An inherent limitation in the current CMS data management model is organization
of data around files. File-based organization fundamentally forces certain data
products, or collections of related fields (data columns) pertaining to an
event to be grouped together with the same data access quality of service
(QoS). In the current workflow model, large-scale tasks transform datasets from
one set of columns to another more compact set, to be able to afford keeping
the latter available with high QoS (i.e., on magnetic disk at several sites.)
It is not currently possible to have different subsets of data columns kept at
high QoS for different datasets at granularities beyond the \emph{data tier}, a
statically defined set of columns. A more flexible data model that can keep
only the needed data columns available at high QoS may significantly reduce
storage volume risk.

\section{Data-Tier Storage Model}

\begin{figure}[htb]
    \centering
    \includegraphics[width=0.8\textwidth,trim={0 2cm 0 5cm},clip]{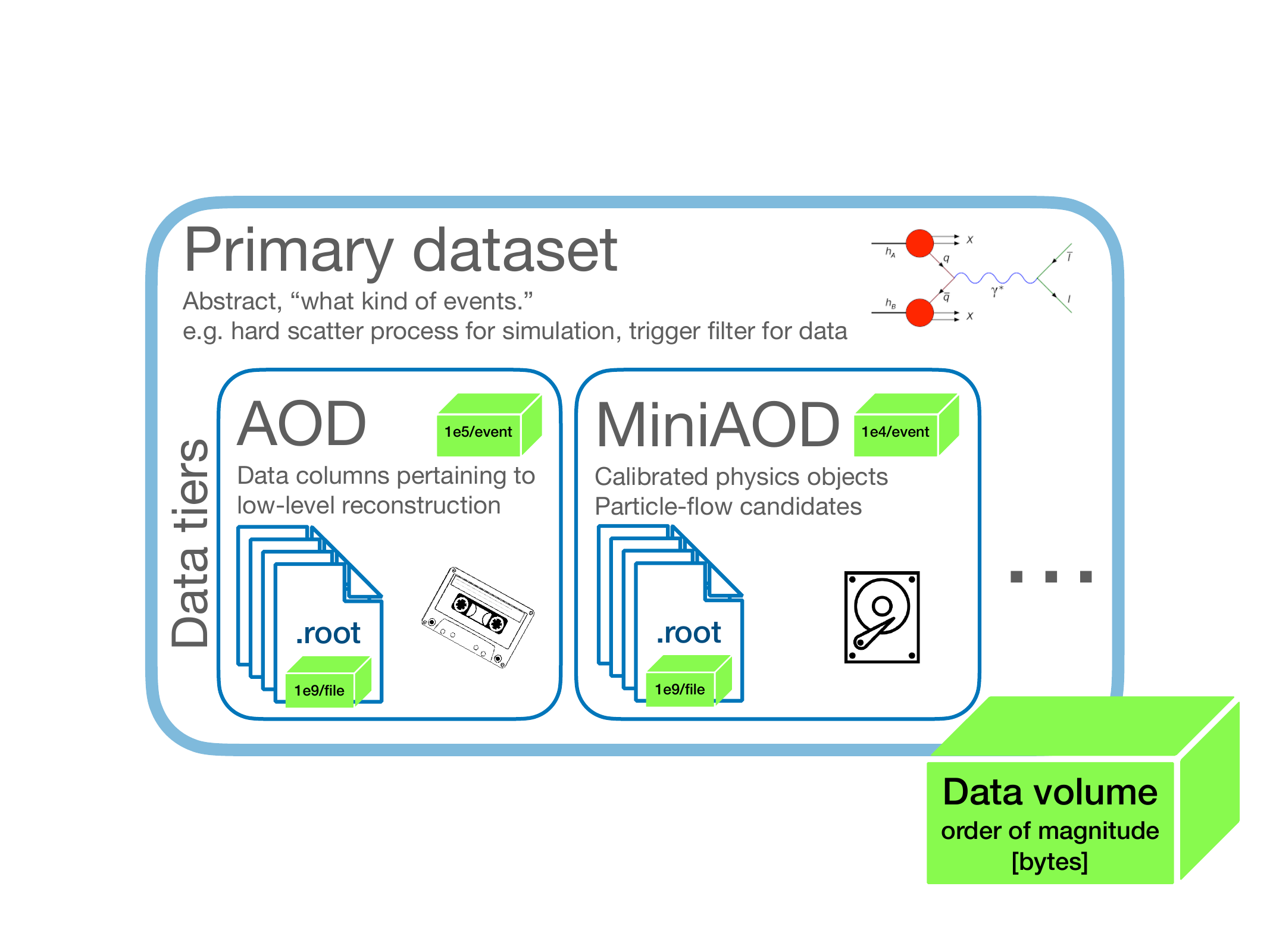}
    \caption{
        Illustration of the data tier model as used for data management in the CMS experiment.
    }
    \label{fig:tiers}
\end{figure}

In the current CMS computing model, \emph{data tiers} represent what subset of
the detector information per event is stored in a particular \emph{dataset}
(e.g.~a collection of files).  The RAW data tier is the unprocessed output of
the CMS detector, and is committed to archival storage (usually tape) while
active storage (usually magnetic disk) is dominated by intermediate
derived/reduced data formats such as AOD and
MiniAOD~\cite{Petrucciani:2015gjw}. For both collider data and simulation,
these reduced data tiers are processed by analysis end-users to form data
formats with simple data types, \emph{NTuples}.  A proliferation of end-user
analysis formats has necessitated frequent access of AOD/MiniAOD tiers up to
the present day. More universal end-user NTuple formats such as
NanoAOD~\cite{Rizzi:2019rsi} effectively define a new data tier and have helped
reduce the need to access upstream data tiers, however they are only expected
to meet the needs of a subset (~50\%) of analysis use cases. An illustration
summarizing the data tier model is shown in Fig.~\ref{fig:tiers}.

Attempts at universal formats such as NanoAOD involve a trade-off in that only
a subset of information is kept for each event. Innovative physics analyses
often require quantities not saved in NanoAOD as they were not considered
pertinent at the time the NanoAOD column set was defined. An analysis that
requires missing columns will have to run their own derivation over MiniAOD or
upstream data tiers to extract the necessary columns and add them to NanoAOD.

Even upstream of NanoAOD formats, the inflexibility of the data tier system
leads to a sub-optimal computing model: analysis users cannot access the AOD
data tier as it is too large to fit the entire CMS working dataset on disk at
that tier. This forces forward-copying certain data products from that tier
onward, duplicating information only for the sake of accessibility; and in the
case of issues in a data re-proecessing (e.g.~re-MiniAOD), the old and new data
may only differ for a small subset of the data columns, yet the remainder have
to be re-written so downstream analysis workflows can access it efficiently.

\begin{figure}[ht]
    \centering
    \includegraphics[width=0.45\textwidth]{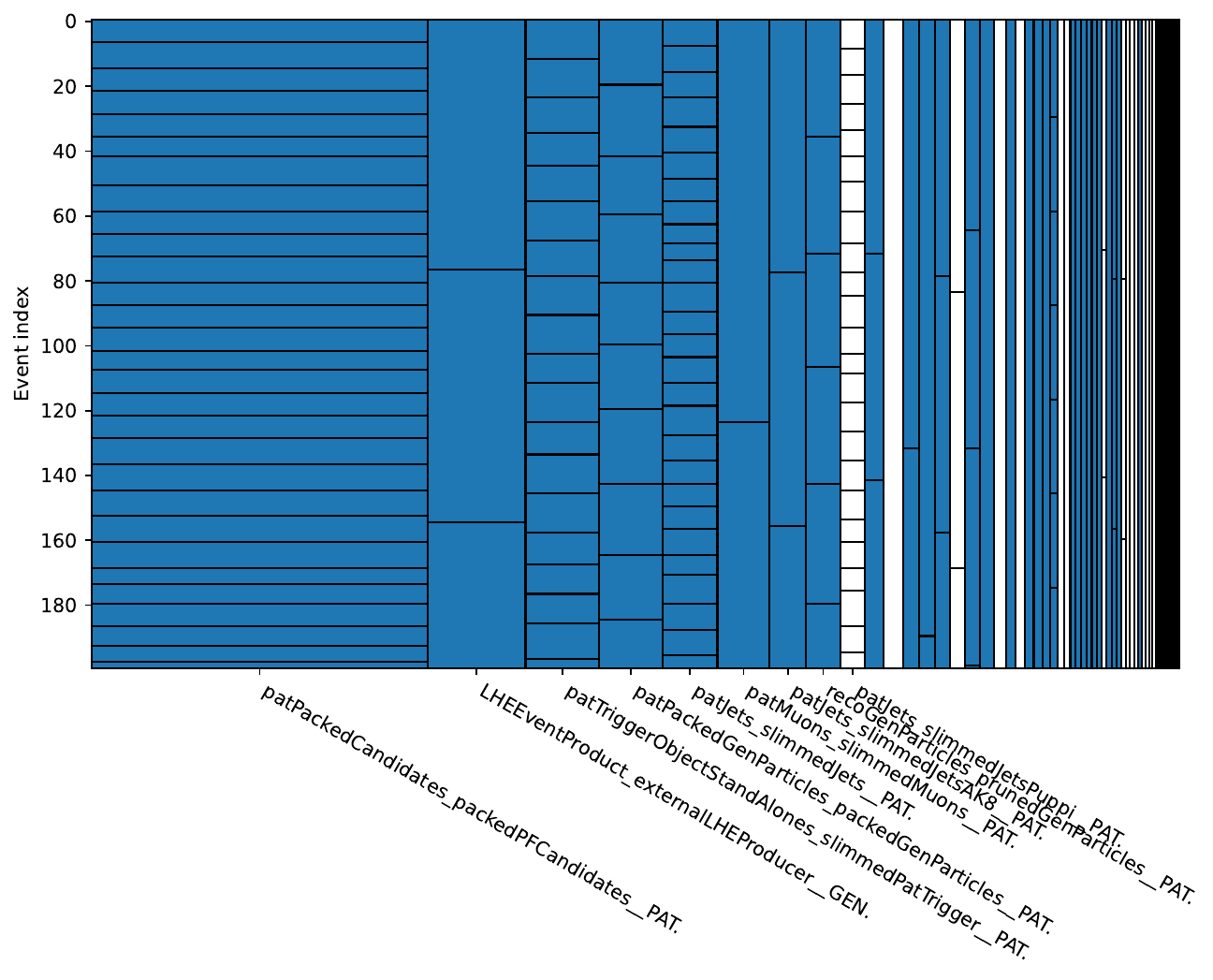}
    \hspace{0.05\textwidth}
    \begin{minipage}[b]{0.45\textwidth}
        \caption{
            Layout of a file in a representative CMS dataset (simulation of semi-leptonic $t\bar{t}$ production),
            where rows represent event indices, and columns are sized proportional to the average compressed size
            of a given data product. The labels of the 9 largest data products
            are shown on the x axis. Unfilled rectangles represent data products that were not accessed.
        }
        \label{fig:tbaskets}
    \end{minipage}
\end{figure}

All centrally managed datasets in CMS are concretely a collection of files
written in the ROOT~\cite{Brun:1997pa} format, with an Events \emph{TTree} and
extra metadata. In this TTree, each data product is stored as a \emph{TBranch}. Some data
product elements are further split into smaller columns for improved
compression, but a product is accessed as one unit within the CMS offline
software (CMSSW) framework~\cite{jones:2006}. TBranch objects reference several
\emph{TBaskets}, each of which contain serialized data products for a range of
events. In Fig.~\ref{fig:tbaskets}, each TBasket accessed by an example CMSSW
workflow reading a MiniAOD file is represented as a rectangle, where the height
is the number of events and width is proportional to the compressed bytes. The
top 6 largest MiniAOD products represent half of the file size. The second
largest data product, \texttt{LHEEventProduct\_externalLHEProducer\_\_GEN.}, is
copied directly from the AOD tier.

\subsection{Projected Usage}

\begin{figure}[ht]
    \centering
    \includegraphics[width=0.45\textwidth]{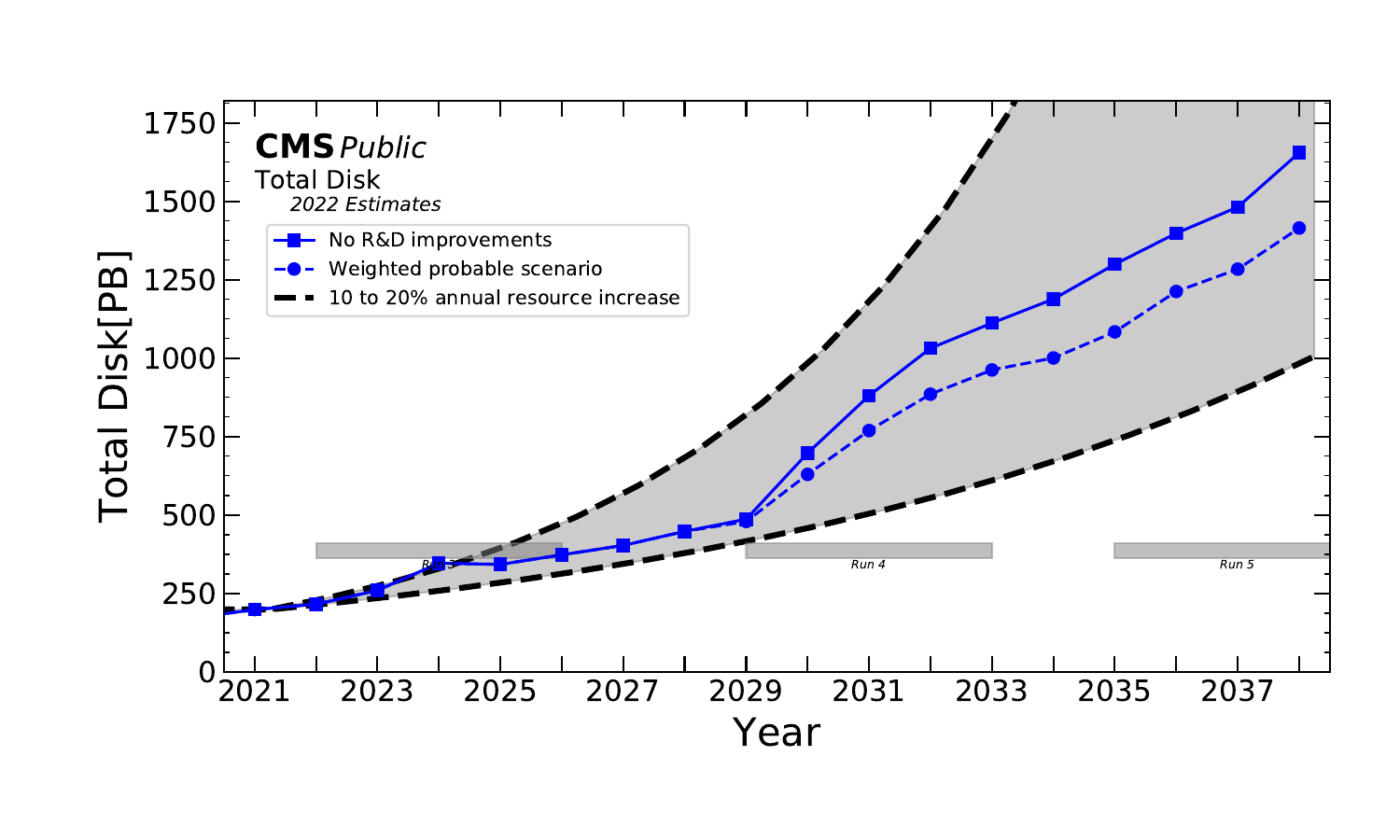}
    \includegraphics[width=0.45\textwidth]{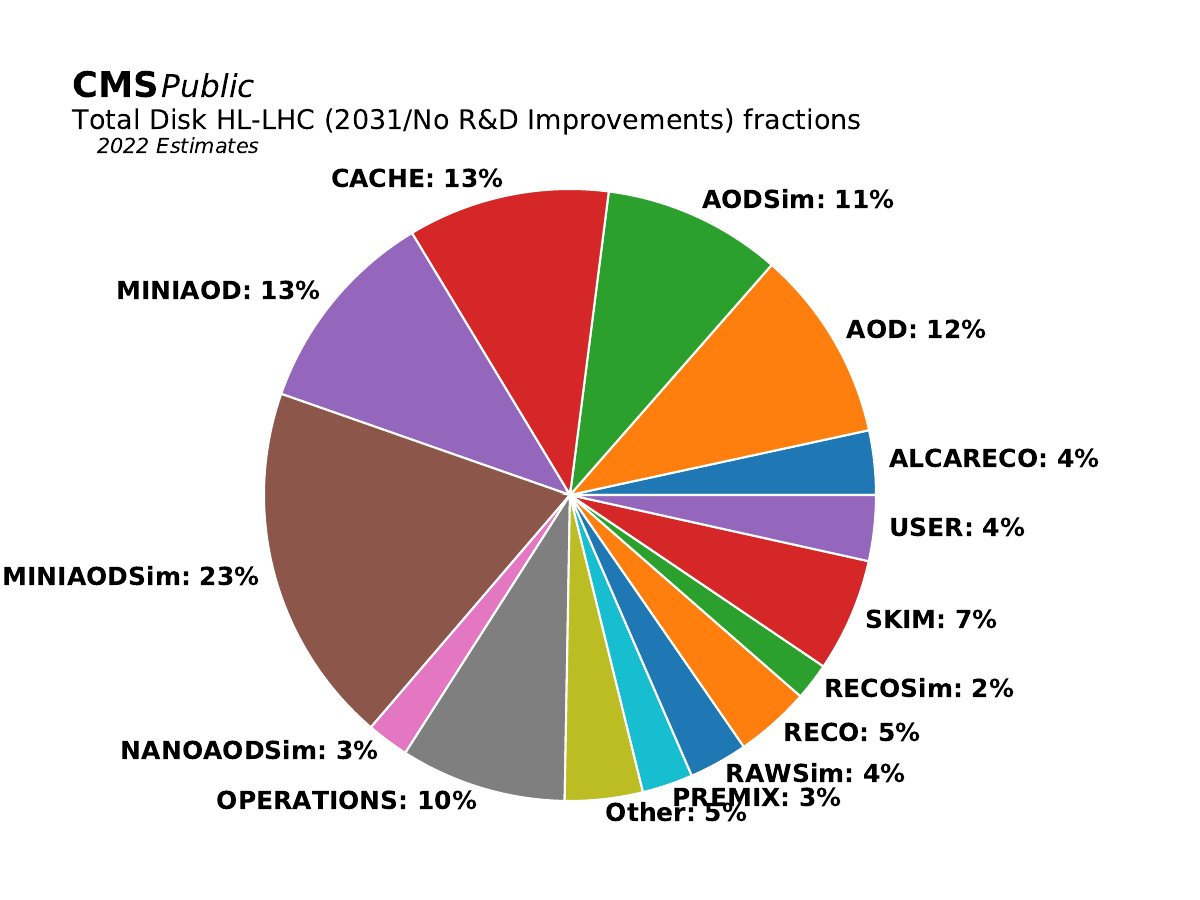}
    \caption{
        Left: annual disk space requirement estimated for CMS processing and analysis needs.
        Right: estimated disk space requirements per data tier for 2031, with a total expected usage of 900 PB.
    }
    \label{fig:cms_disk_proj}
\end{figure}

CMS computing plans for the HL-LHC assume that a majority of analyses will
utilize centrally produced NanoAOD and that the bulk of intermediate formats
currently kept on disk (often with multiple copies) will not be kept on active
storage. Even with such assumptions, the disk storage needs of CMS in the first
year of Run 4 will exceed one Exabyte across all sites, representing a factor
of four increase over the needs in Run 3 (Fig.~\ref{fig:cms_disk_proj},
left)~\cite{cms-oandc-public}. A plethora of non-NanoAOD data tiers
(Fig.~\ref{fig:cms_disk_proj}, right) is expected to be available on disk in
2031 in current projections, and the mixture may well change with detector
commissioning needs. 

\section{Object Stores}
\label{sec:object}

Object stores, which operate on a key-value principle, allow for
highly-granular access to information via efficient metadata lookup. The
evaluation of sub-file or object-based granularity is a key R\&D goal for Data
Organization Access and Management (DOMA) in the HEP Software Foundation
Community White Paper~\cite{hsf-whitepaper} which was further enforced in the
2021 Snowmass Planning Process~\cite{bhimji2022snowmass}. Object stores are
widespread in industry and are near-universal in cloud storage with the Amazon
S3 API emerging as a {\it de facto} access protocol across them. A central
object store can also provide the backbone of a content delivery network (CDN)
for end-users and allow for analysis without need for an analysis NTuple format.

To concretely explore the possibilities of object storage for event data in
CMS, we developed an object data format, a test cluster using Ceph object store
technology, and a S3 input and output module in a prototype event processing
framework similar to CMSSW, as discussed in the following subsections.

\subsection{An object data format}
\label{sec:objformat}

In our object data format, one index object holds metadata for an entire
primary dataset. For each data product, or column of data, a \emph{stripe} of
events is written as an object in an S3 bucket. Each event data product is
serialized with standard ROOT IO. The stripe size is chosen on first write,
once the compressed output buffer reaches a \emph{target stripe size}, usually
100k-1MB, and the number of events evenly divides a configurable \emph{event
batch size}. The index object may be replaced by a database, to more easily add
and remove products. By forcing the number of events to be a constant even
divisior of the event batch size, the metadata volume grows as (N products) +
(M events) despite the N*M growth of the number of stripe objects. The data
stripes will not exactly meet the target size, but will have a distribution
centered very close to the target due to the independent
identically-distributed nature of the event data.  To handle very small data
products which would never reach the target stripe size, we optionally collect
them into product groups following a greedy algorithm. This does not inherently
prevent independent access of columns in that stripe, as byte-range addressing
is available in the S3 protocol. Although this object store system assumes all
objects are online-accessible, infrequently used column stripes could be
concatenated and offloaded to tape systems based on a caching policy.

Significant improvements in long-term disk storage needs can be realized with
this object store scheme. In a mock example (see Table \ref{tab:mock}), using
average per-data-product sizes for a CMS semi-leptonic $t\bar{t}$ production primary
dataset, two MiniAOD versions (v1, v2) are produced from the same parent
dataset, where the latter is produced to update electron data. In the object
store scheme, there is no need to re-produce other (unchanged) data products.
In any case, data products such as \texttt{genParticles} would never need
to be copied forward in the MiniAOD production.

\begin{table}[htb]
    \centering
    \caption{Mock data usage example}
    \label{tab:mock}
    \begin{tabular}{lcc}
        \hline
        Data product & \multicolumn{2}{c}{kilobytes per event} \\
                    & Data tier model & Object store model \\
        \hline
        genParticles (upstream) & 5.7 & 5.7 \\
        genParticles (MiniAODv1) & 5.7 & - \\
        genParticles (MiniAODv2) & 5.7 & - \\
        slimmedElectrons (MiniAODv1) & 1.3 & 1.3 \\
        slimmedElectrons (MiniAODv2) & 1.3 & 1.3 \\
        Other event products (MiniAODv1) & 48.7 & 48.7 \\
        Other event products (MiniAODv2) & 48.7 & - \\
        \hline
        Total & 117.1 & 57
    \end{tabular}
\end{table}

\subsection{Test Ceph Cluster}
\label{sec:cluster}

A pilot cluster has been assembled to gain experience in managing a Ceph
installation and for evaluation purposes. Nine disk servers, retired from the
Fermilab dCache disk pool, provide a total 2 PB of disk drives of vintage
2014-2018, totaling 288 Ceph Object Store Daemons (OSDs).  Two servers provide
20 TB of NVMe across 32 OSDs for the metadata pool. All machines have 10 to
100Gbps networking. The cluster supports filesystem access using the xrootd~\cite{xrootd}
protocol via an edge machine with CephFS mounted, and S3 object access via a
RadosGW edge machine. For all performance tests, client machines are accessing
the cluster from within the Fermilab network.

For the S3 object access, three buckets are set up with different Ceph pool
configurations.  In the first configuration, each object is split into 16 kibibyte (KiB)
chunks and each chunk is further split into 4 data blocks of 4 KiB and 2 parity
blocks using an erasure coding scheme, dubbed \emph{EC4+2}. The bucket also
maintains an index of all objects in it, where each item in the index contains
about 300 bytes of metadata such as the last access time and owner. The index
is stored in a triply-replicated NVMe pool using Ceph's key-value object type.
In a second configuration, the same erasure-coded pool is used but the bucket
index is disabled. Without the bucket index, objects can still be read and
written, but cannot be listed, so an external metadata database would be
necessary to manage object lifecycle properly.  In the last configuration, all
data is stored a triple-replicated disk pool, dubbed \emph{Rep3}.

\subsection{Prototype Client}

The HEP-CCE \texttt{root\_serialization} project provides a C++ framework for
performance experiments with various I/O packages from within a multi-threaded
program.~\cite{root-serialization} The program mimics behaviors common for HEP data processing
frameworks, and uses a Intel Thread Building Blocks (TBB) thread pool to
schedule parallel \emph{tasks}. The number of threads in the TBB thread pool is
configurable. Within this framework, S3 input and output modules have been
developed to read from and write to S3 buckets, respectively. This allows to
explore the performance and parallelization behavior of writing to an Ceph S3
service in comparison to local files with ROOT or other formats. The S3
protocol is implemented with libs3~\cite{libs3}. Requests are performed
asynchronously in a separate thread from all other tasks. Network errors are
handled with an exponential backoff with retry mechanism.

%

\section{Performance Results}

Using the object data format, test cluster, and prototype event processing
framework described in Section \ref{sec:object}, we evaluated the storage and
read/write performance against a baseline of current MiniAOD event size and
processing rates.  Scaling behavior is probed both in single- and multi-process
setting, to understand threading efficiency and client load limits of the
cluster, respectively.

\subsection{Storage Efficiency}

\begin{figure}[htb]
    \centering
    \includegraphics[width=0.45\textwidth]{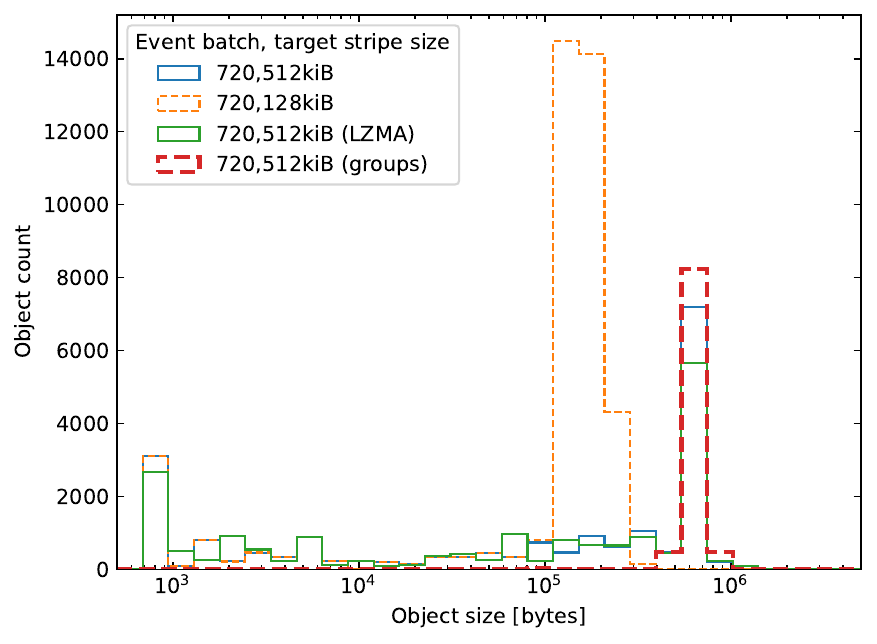}
    \caption{
        Distribution of object stripe sizes in the output dataset, for
        various configurations of the object data format writer.
    }
    \label{fig:objectsizes}
\end{figure}

A MiniAOD input dataset is converted into the object data format with various
settings of event batch size, target stripe size (Section \ref{sec:objformat}),
and stripe compression. The storage efficiency, measured in average kilobytes
per event, is shown in Table \ref{tab:storage}. We find that the object storage
format has generally a larger size per event than the input, especially when
LZMA compression is not used.  Even when LZMA is used, the object format cannot
compress as well as MiniAOD due to a limitation in the ROOT object
serialization when not used in conjunction wtih TTree I/O: the object fields
cannot be split further into struct-of-array types, which hinders
compressability.  A special consideration for Ceph is that small objects incur
additional storage overhead (listed in \% ) due to the minimum object size
granularity of 4 KiB. With larger target stripe sizes, as well as with use of
product groups, this overhead is reduced. In Fig.~\ref{fig:objectsizes}, the
distribution of output stripe sizes is shown for each of the storage
configurations shown in Table \ref{tab:storage}. The amount of small product
stripes is significantly reduced with the use of product groups.

\begin{table}[htb]
    \centering
    \caption{Comparison of data volume with object store scheme and input MiniAOD ROOT file}
    \label{tab:storage}
    \resizebox{\textwidth}{!}{
    \begin{tabular}{lcccc}
        \hline
        Format	& Compression & Event batch size & Target stripe size & kB per event \\
        & & & & (granularity overhead) \\
        \hline
        MiniAOD input	& LZMA & - & - & 55.7 \\
        Object  & ZSTD & 720 & 128KiB & 71.4 (6.5\%) \\
        Object  & ZSTD & 720 & 512KiB & 70.6 (3.5\%) \\
        Object  & LZMA & 720 & 512KiB & 61.8 (3.7\%) \\
        Object (groups)  & ZSTD & 720 & 512KiB & 70.6 (1.4\%) \\
        \hline
    \end{tabular}
    }
\end{table}

\subsection{Single-client scaling}
\label{sec:singleclient}

\begin{figure}[htb]
    \centering
    \includegraphics[width=0.45\textwidth]{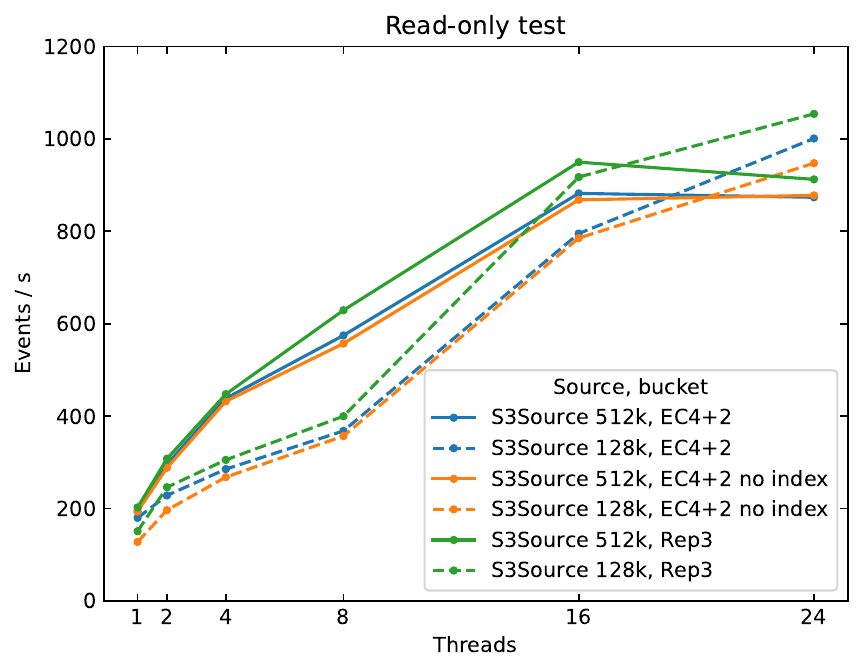}
    \includegraphics[width=0.45\textwidth]{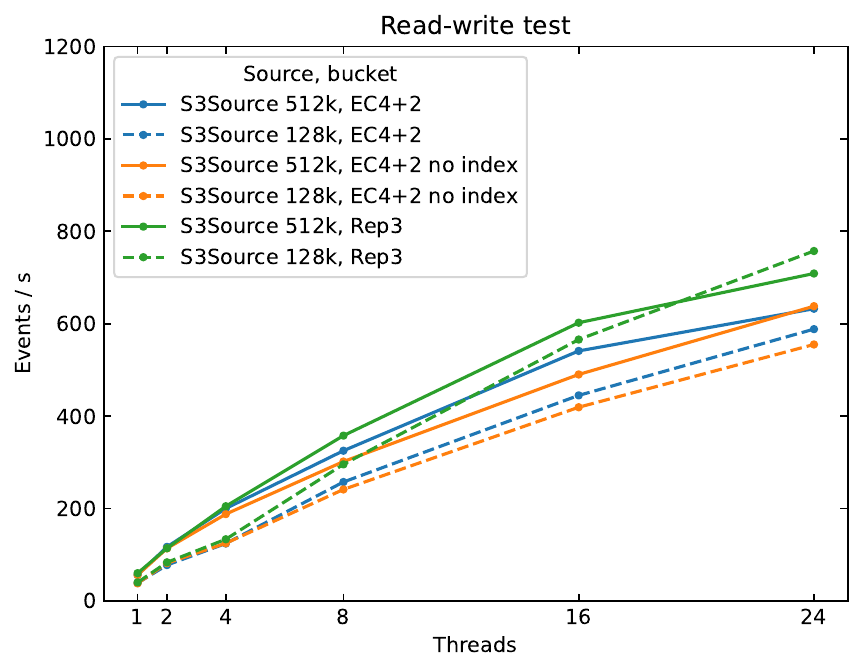}
    \caption{
        Single-client read-only (left) and read-write (right) thread scaling tests.
    }
    \label{fig:multiclient-bandwidth}
\end{figure}

We performed a set of tests where all data products in the MiniAOD tier are
read and written as fast as possible to/from the test object store cluster,
using a single executable client process. The executable’s thread scaling
properties are probed, with the metric being the events processed per second.
In the read-write test, events are: read, decompressed, deserialized,
serialized, compressed, and written. For the read-only chain, only the first
three steps are performed. Tests were run on a 24-core machine with 10Gbps
network connection to the Ceph cluster. The tests are performed for two
configurations of event batch size and target stripe size. For each
configuration, the source and output modules target one of three S3 buckets
with different storage configurations, as described in Section
\ref{sec:cluster}. Fig.~\ref{fig:multiclient-bandwidth} shows good scaling
behavior with increased thread count, and negligible performance difference
between the storage configurations. As a point of reference, CMS production
jobs reading MiniAOD and producing NanoAOD run at a CPU-limited rate of about
10 events/second/thread.

\subsection{Multi-client scaling}

\begin{figure}[htb]
    \centering
    \includegraphics[width=0.45\textwidth]{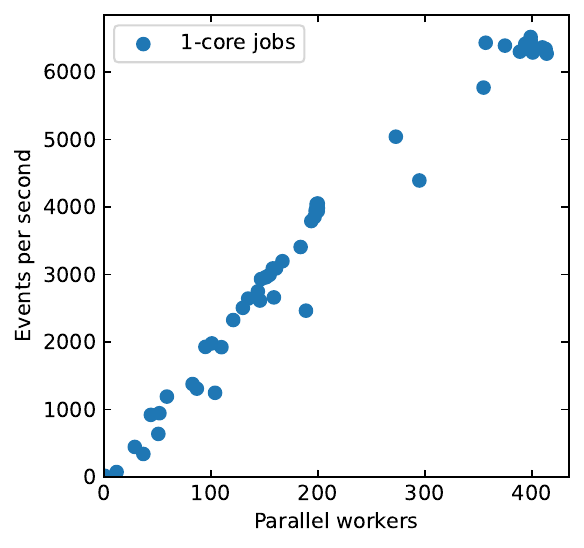}
    \includegraphics[width=0.45\textwidth]{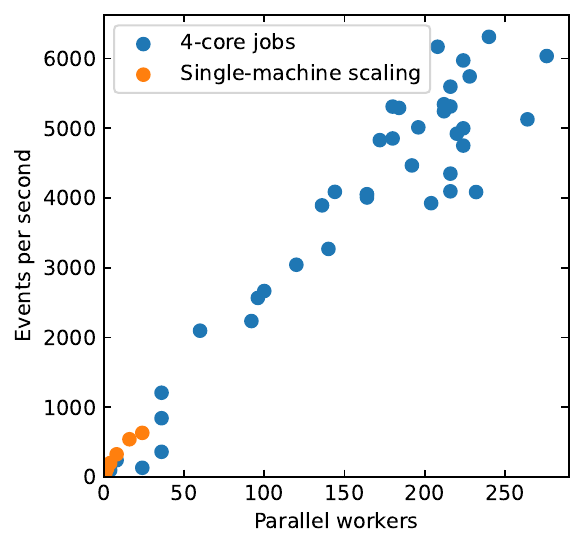}
    \caption{
        Scaling
        Left: multi-client write test.
        Right: read-only test.
    }
    \label{fig:multiclient-scaling}
\end{figure}

We performed a test of converting data in MiniAOD from the ROOT file format to
the object format at scale, where the input files are read from Fermilab's
dCache disk cluster over xrootd and written to the test cluster via S3.  In
this test, the conversion processes are single-threaded, and up to 400
simultaneous workers are writing objects with a target size of 512KiB to the
EC4+2 bucket. The events per second written to the object store is shown as a
function of worker count in Fig.~\ref{fig:multiclient-scaling}, left. A saturation
point is observed with approximately 350-400 workers writing 6300 events/s (450
MB/s) to the data pool. In total, 4.5 TB of data was written into 7.4 million
objects.

Once the data was written, we performed a read-only test similar to Section
\ref{sec:singleclient}, with 4 threads per worker and several independent
worker processes accessing the test cluster in parallel. The results are shown
in Fig.~\ref{fig:multiclient-scaling}, right.  In this case, there is a larger
variation in cluster performance at high worker count, and no apparent
saturation is reached. Further tests with additional workers are necessary. In
this test, we can directly compare the scaling performance with the
single-client scaling results, and they appear consistent.

While the scaling tests were in progress, performance profiles of the client
application showed that the I/O latency is fully hidden when the S3 server is
not heavily utilized, but server saturation can cause the client applications
to stall and show poor CPU efficiency. As this is a client application with no
CPU-heavy tasks other than decompression and deserialization, we expect that in
a realistic processing task, the I/O latency would remain in the shadow of the
CPU-bound TBB tasks.

\section{Conclusion}

In conclusion, the object data format provides novel data management capabilities
with respect to a data tier and file-based format. In particular, it shows promise
in reducing the total storage requirements as well as providing more flexibility
in defining what data are easily accessible to analysis tasks. In a prototype
processing framework accessing a Ceph S3 object storage cluster, we find that the
data volume is in line with expectations and service scaling is promising, with one
Ceph RadosGW node serving 350-400 parallel clients before performance saturation.

To fully realize the possibilities of this object data format, additional
software development will be needed. A high priority is to migrate the object
I/O modules from the prototype event processing framework to CMS offline
software, possibly leveraging the RNtuple integration efforts, as RNtuple also
provides an object backend with a very similar design as the one presented
here.~\cite{rntupleobj} A new metadata service will be required to track which
objects are available, and close integration with the workflow management
system will be required.

\section*{Acknowledgements}

This work is supported by the US-CMS HL-LHC R\&D initiative, and Fermi Research
Alliance, LLC under Contract No.~DE-AC02-07CH11359 with the U.S.~Department of
Energy, Office of Science, Office of High Energy Physics.

\bibliography{ncsmith-chep2023-s3.bib}

\end{document}